\documentclass[pre,twocolumn]{revtex4}
\usepackage{graphicx}

\newcommand{\diff}{{\mathrm d}}

\begin{document}

\title{Failure of deterministic and stochastic thermostats
to control temperature of molecular systems
}
\title{Failure of deterministic and stochastic thermostats
to control temperature of molecular systems
}

\author{Hiroshi Watanabe}
\email{hwatanabe@issp.u-tokyo.ac.jp}
\thanks{Corresponding author}
\affiliation{
The Institute for Solid State Physics, The University of Tokyo,
Kashiwanoha 5-1-5, Kashiwa, Chiba 277-8581, Japan
}

\begin{abstract}
We investigate the ergodicity and ``hot solvent/cold solute" problems in
molecular dynamics simulations. While the kinetic moments and the stimulated Nos\'e--Hoover methods improve the ergodicity of a harmonic-oscillator system, both methods exhibit the ``hot solvent/cold solute" problem in a binary liquid system. These results show that the devices to improve the ergodicity do not resolve the ``hot solvent/cold solute" problem.
\end{abstract}

\maketitle

Since Alder and Wainwright performed molecular dynamics (MD) simulations for the first time~\cite{Alder1957},
they have been useful tools for exploring a wide variety of science.
Traditional MD simulations lead to an isoenergetic condition, that is,
the total energy of the system is conserved throughout its time evolution.
However, the temperature dependence of observables is usually of interest, instead of their total energy dependence. For this purpose, the Nos\'e--Hoover thermostat is widely used
since this method generates the canonical distribution for all degree of freedom by the simple modification of equations of motion~\cite{Hoover1985}.
However, it has been found that the Nos\'e--Hoover method sometimes fails to the control temperature.
One is the ergodicity problem of the harmonic oscillator~\cite{Hoover1985}.
A harmonic oscillator with the Nos\'e--Hoover method loses ergodicity owing to a nontrivial conserved value, and therefore, the distribution deviates from the canonical distribution~\cite{Watanabe2007, Legoll2007}.
To address this problem, multivariable thermostats have been proposed~\cite{Martyna_1992, Ju_1993, Hoover_1996}.
Recently, Hoover \textit{et al.} proposed a new method that achieves ergodicity in a harmonic oscillator with only a single variable~\cite{Hoover_2016}.
In addition to the ergodicity problem, another type of problem exists.
When a thermostat is attached to an inhomogeneous solute--solvent system,
the solute and solvent are equilibrated at the different temperatures.
This problem is referred to the ``hot-solvent/cold-solute" problem~\cite{Lingenheil2008}. 
The purpose of the present manuscript is to clarify these two problems, the ergodicity and ``hot-solvent/cold-solute" problems, result from different causes.

\begin{figure}
\centering
\includegraphics[width=6.5cm]{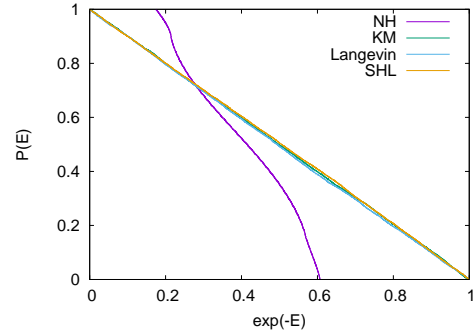}
\caption{Cumulative distribution functions.
The results of the Nos\'e--Hoover (NH), kinetic moments (KM), Langevin, and stimulated Nos\'e--Hoover  (SHL) methods are shown. 
The fictitious masses of thermostats are set to unity.
The fourth-order Runge--Kutta method is used for the deterministic thermostats
and the first-order Euler method is used for the stochastic thermostats.
The initial condition is $\{p, q\} = \{0, 1\}$ and the additional variables of the 
thermostats are set to zero.
}
\label{fig_cdf}
\end{figure}

We consider four thermostats, 
Nos\'e--Hoover, 
kinetic moments~\cite{Hoover_1996},
Langevin~\cite{Allen_1989}, and 
stimulated Nos\'e--Hoover methods~\cite{Samoletov2007}.
The Nos\'e--Hoover method fails to control the temperature of a harmonic oscillator.
To address the ergodicity problem, the kinetic moments method is proposed~\cite{Hoover_1996}.
Since the kinetic moments method has two or more additional variables, the system does not have any nontrivial conserved value, and therefore, the ergodicity is recovered.
The Langevin method, which is a stochastic thermostat, can also achieve the canonical distribution when it is attached to a harmonic oscillator.
While all degree of freedoms are subject to stochastic control in the Langevin method, 
Samoletov \textit{et al.} proposed a new thermostat, the stimulated Nos\'e--Hoover method~\cite{Samoletov2007}.
In the stimulated Nos\'e--Hoover method, only the additional variable of the Nos\'e--Hoover method,
which is usually denoted by $\zeta$, is subject to stochastic perturbation.
This method ensures the ergodicity.
Figure~\ref{fig_cdf} shows whether four thermostats control the temperature correctly.
Let $P(E)$ be the cumulative distribution function (CDF) of the energy $E$,
\textit{i.e.}, the probability that the energy of the system is less than $E$.
If the system is in equilibrium with temperature $T$, then $P(E) = 1 - \exp(E/T)$.
Here, the Boltzmann constant is set to unity.
It is showed that $P(E)$ significantly deviates from
the expected distribution, and therefore, the Nos\'e--Hoover method fails to control the temperature of this system, while the other three achieve the canonical distribution correctly.

\begin{figure}
\centering
\includegraphics[width=6.5cm]{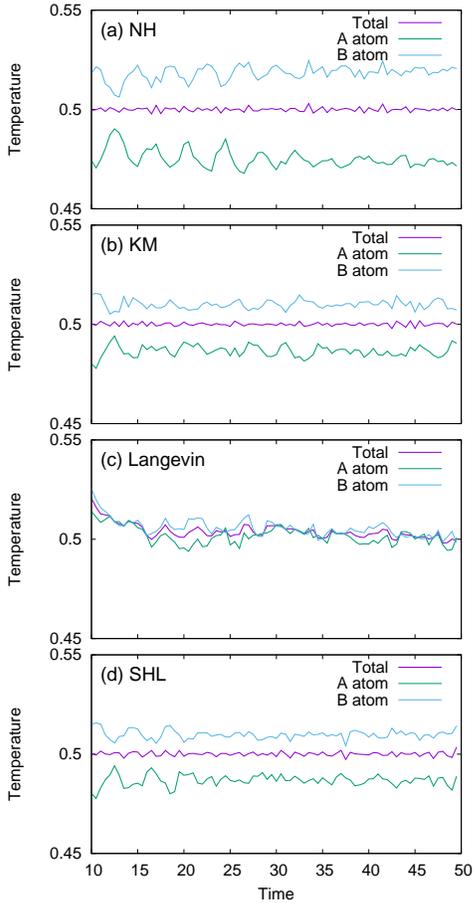}
\caption{
Time evolutions of the temperatures of the total system,
$A$-atoms, and $B$-atoms.
Four types of thermostats are attached, the
(a) Nos\'e--Hoover (NH),
(b) kinetic moments (KM),
(c) Langevin, and
(d) stimulated Nos\'e--Hoover (SNH) methods.
}
\label{fig_bindroplet}
\end{figure}

Next we consider the ``hot solvent/cold solute" problem.
While this problem is frequently found in solvent--solute systems, 
we consider a simpler system, a binary-liquid system.
This system consists of two types of atom, $A$ and $B$, referred to as $A$- and $B$-atoms.
All atoms have an identical mass and diameter.
We consider the equimolar case, that is, the system consists of 
the same number of both types of atom.
The interaction between atoms of the same kind is given by the truncated Lennard-Jones (LJ) potential
and the interaction between different types of atoms is given by the repulsive Weeks--Chandler--Andersen (WCA) potential~\cite{Weeks_1971}.
We choose the truncation length to $3 \sigma$, where $\sigma$ is the diameter of atoms.
Hereinafter, we measure observables in LJ units, that is, the length is measured by $\sigma$, and so forth.
While atoms of same type, \textit{i.e.}, $A-A$ and $B-B$, have an attractive interaction,
different types of atoms only have a repulsive interaction.
Therefore, the system exhibits phase separation at a sufficiently low temperature.

The atoms are setup in the face-centered cubic lattice, and random velocities are assigned to the atoms with an identical absolute value.
The simulation box is a cube of length $L=30$.
The total number of atoms is 16,384.
The isothermal simulations are performed using thermostats with the desired temperature $T_0 = 0.5$.
The above conditions are chosen so that the system exhibits phase separation
and both atoms are in liquid phase.
The two types of atom are placed so that the system has a spherical interface,
with the $A$-atoms located inside and the $B$-atoms located outside.
The time step is 0.005.

The time evolutions of temperatures are shown in Fig.~\ref{fig_bindroplet}.
We observed the temperature of the total system,
that of the $A$-atoms, and that of the $B$-atoms.
There are significant differences between the three types of temperature when 
the Nos\'e--Hoover, kinetic moments, and stimulated Nos\'e--Hoover methods are used.
This difference is caused by the phase separation.
When $A$-atoms form a droplet, their temperature decreases,
while the temperature of $B$-atoms increases.
The difference in the temperatures of the two subsystems remains for long time
since the thermostats cannot control the temperature of each subsystem independently.
When we attach the Langevin method to the system, the temperature of all atoms are controlled independently,
and consequently, there is no significant difference between the three types of temperature as shown in Fig.~\ref{fig_bindroplet}~(c).
This problem occurs when we adopt global thermostatting, \textit{i.e.}, only one thermostat 
controls the temperature of the system. 


\begin{table}
\begin{tabular}{cccc}
\hline
Method & Time evolution & Harmonic oscillator & Binary liquid \\
\hline
NH & Deterministic & NG & NG \\
KM & Deterministic & OK & NG \\
Langevin & Stochastic & OK & OK \\
SNH & Stochastic & OK & NG \\
\hline
\end{tabular}
\caption{
The successes or failures of thermostats in controlling temperature 
of harmonic oscillators and binary-liquid systems.
'NG' denotes that a thermostat cannot provide the
canonical distribution for a harmonic oscillator
and that a thermostat thermalizes each component of liquid
at different temperature for a binary-liquid system.
}
\label{table_thermostats}
\end{table}

In the present short notes, we investigated the properties of four types of thermostat,
the Nos\'e--Hoover,
kinetic moments,
Langevin, and stimulated Nos\'e--Hoover methods.
The former two are deterministic, while the latter two are stochastic.
We considered two types of system, a harmonic oscillator
and a binary liquid. The success or failure in controlling the temperature
is summarized in Table~\ref{table_thermostats}.
The ergodicity problem occurs when there exists a nontrivial conserved value.
This problem can be avoided by adding degree of freedom to a thermostat or
using stochastic thermostats.
The ``hot solvent/cold solute" problems occurs when a system is singly thermostated.
This problem is nothing to do with the fact that a thermostat has many degrees of freedom
or a thermostat is stochastic. Therefore, the devices to improve the ergodicity
of harmonic oscillator do not resolve the `hot solvent/cold solute" problem.

\section*{Acknowledgements}
This work was supported by JSPS KAKENHI Grant Numbers 15K05201 and
by MEXT as ``Exploratory Challenge on Post-K computer" (Frontiers of Basic Science: Challenging the Limits).

\bibliography{itherm}

\end{document}